# INTERMODAL COMPETITION IN THE BRAZILIAN INTERSTATE TRAVEL MARKET****

## FREDERICO ARAUJO TUROLLA*
Escola Superior de Propaganda e Marketing and Fundação Getúlio Vargas

## MOISES DINIZ VASSALLO**
Núcleo de Estudos em Competição e Regulação do Transporte Aéreo,
Instituto Tecnológico de Aeronáutica

## ALESSANDRO VINICIUS MARQUES DE OLIVEIRA***
Núcleo de Estudos em Competição e Regulação do Transporte Aéreo,
Instituto Tecnológico de Aeronáutica

**Abstract**

*This paper presents a test of intermodal interaction between coaches and airlines in Brazil in order to check for the efficacy of recent liberalization measures designed to promote competition in both industries. Interstate travel service in the country is heavily provided by coaches, and the system is fully operated by the private sector under public delegation through permits and authorizations. Agency-based regulation was introduced in 2002 along with a price cap regime aimed at enhancing the flexibility to change fares in response to demand and cost conditions. By making use of a reaction function-based model of coach operators' pricing decisions in the interstate travel market, we then estimate the sensitivity of the changes in coach fares to the changes in airline fares in a simultaneous-equation framework. Intermodal interaction among coach operators and airlines is*

* Escola Superior de Propaganda e Marketing (ESPM) and Fundação Getúlio Vargas (FGV). R. Dr. Neto de Araujo 320 cj 1307, Vila Mariana, São Paulo, Brazil. E-mail: fredturolla@pezco.com.br.
** Center for Studies of Airline Regulation and Competition, Aeronauthical Institute of Technology (NECTAR/ITA). Av. Escola Politécnica 2.200, AP 73, Bloco Itália - Jaquaré, São Paulo, Brazil. E-mail: vasallo@fipe.org.br
*** Center for Studies of Airline Regulation and Competition, Aeronauthical Institute of Technology (NECTAR/ITA). Rua Vera Lúcia Cardoso Silva Barros 84, Urbanova, São José dos Campos, Brazil. E-mail: A.V.M.Oliveira@gmail.com
**** The author wishes to thank Manuel Willington, Claudio Agostini, the State of São Paulo Research Foundation (FAPESP) and the National Council for Scientific and Technological Development (CNPq).

*found to be highly significant and probably due to the competition for a small but increasing set of premium, quality-sensitive, coach passengers.*

Keywords: *Air Transportation, Coaches, Industrial Organization, Econometrics.*

JEL Classification: *L90, C33.*

## I.   Introduction

This paper aims at developing an empirical model of pricing decisions for the interstate travel market in Brazil, in order to investigate potential intermodal competition. Interstate travel service within the country is heavily provided by coaches, which account for roughly ninety five percent of passengers; travel by plane is nowadays a potential substitute, although air tickets –usually much more expensive relatively to coach travel– are still unaffordable to the majority of coach travelers. We expect that intermodal interaction among coach operators and airlines may emerge as a result of a competition for a small but increasing set of premium, quality-sensitive, coach passengers, who may be willing to pay for air travel during airline fare wars.

Intermodal interaction may be regarded as a relevant issue nowadays, and yet to be further investigated by the literature; notable examples are Wardman (1997), and Ivaldi and Vibes (2005). In fact, Wardman (1997) points out that "*relatively little is known about the interaction between modes in the inter-urban travel market*".

The system of interstate travel by coach in Brazil is fully operated by the private sector under public delegation, through permits and authorizations. Agency-based regulation was introduced in 2002 and this new regulatory framework may allow for further advances in competition. A price cap regime was introduced in order to enhance competition among operators, although it is still not possible to set prices below the tariff ceiling without previous communication to the agency.

By analyzing pricing decisions of operators in some cities within the country, here we intend to investigate potential price sensitivity to cost components and competitive pressure from airlines; the latter became certainly more relevant after the liberalization of the airline industry, during the nineties, and the recent entry of the first low cost airline of Latin America, *Gol Linhas Aéreas*, in January 2001 (Oliveira, 2008).

The paper is organized in five parts. Section II discusses the relevant features of the interstate travel market in Brazil; Section III presents the theoretical framework of intermodal competition; Section IV develops the empirical model of pricing decisions by coach operators along with estimation results; and finally, conclusions are presented.

## II.   The Interstate Travel Market in Brazil

Road transportation is a key sector of the Brazilian economy. Since the late fifties, when the federal government adopted the strategy of boosting and promoting the development of the local automobile industry, the majority of people and cargo in the interstate travel market has been extensively carried on highways. Brazil has 1.8 million kilometers of highways, but only ten percent of these are paved. Of the total, approximately sixty thousand kilometers are paved federal highways[1].

Figure 1 presents a map with the main federal and state highways within the country.

Although the national highway system is the second of the world in size[3], it is traditionally marked by public and private underinvestment with perverse effects on the quality of the service provided by carriers. Negative spillovers of the lack of investments in road infrastructure include not only the comfort and safety of passengers but also the logistics of national corporations and the competitiveness of exports. In fact, a relevant subset of the system has historically been in very bad conditions and thus constraining the growth of the interstate travel market. This situation, along with the recent economic liberalization measures of the domestic airline industry and the consequent fall in air fares, have been stimulating travelers to increasingly consider air transportation in their choice set.

Historically, the strict regulation of the road transportation market led to a virtual absence of competition between carriers. The regulatory framework not only inhibited new entry and sustained a high concentration level but also prevented incumbents to engage in price competition. This situation led to a significant lack of efficiency and low productivity until recently, in such a way that some analysts used to refer to the sector as dominated by a "cartel". Also, until recently, intermodal competition between coach and planes was very limited.

## FIGURE 1

MAIN FEDERAL AND STATE HIGHWAYS IN BRAZIL (2000)[2]

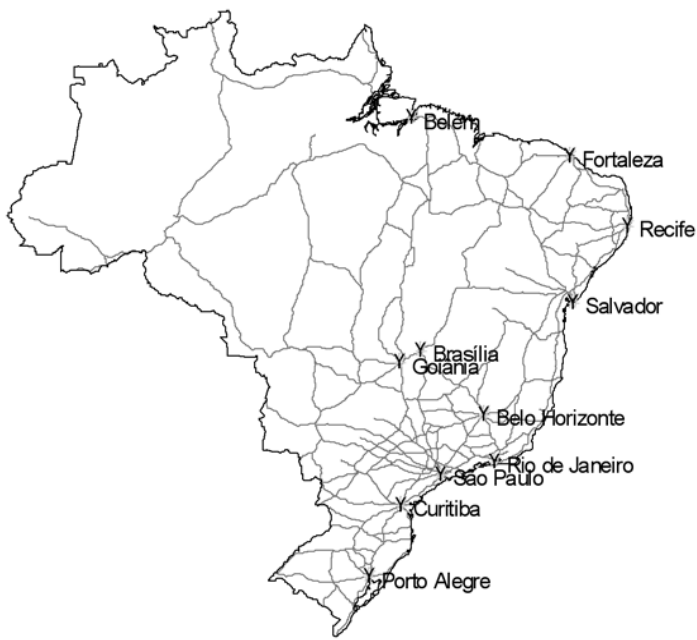

Competition by passengers from other modes is relevant only in very specific markets. The Brazilian train system is under-developed and the amount of passengers carried in interstate routes is overall negligible, except for the regular routes operated by only two out of twelve private railways. These two carriers (EFVM and EFC) carry only 1.5 million passengers a year[4], less than 1% of the total coach passengers. Waterways play some role in some states but are also very specific to those regional markets. Finally, "self transportation" can be seen as a substitute to passenger transportation services by road, as supported by Guimarães and Salgado (2003).

In 2006, there were 222 interstate coach operators in Brazil, with a fleet of 15.616 buses. The biggest individual operator, Viação Itapemirim, had a 12.2% share under the passengers times kilometers criterion and none of the remaining operators held more than 6.4% of the market. The Herfindahl-Hirschman Index (HHI), calculated under the assumption of individual operators being independent companies, was 356 and the CR4[5] was 0.294. However, actual concentration in relevant markets may be quite different from this general calculation, since: i) relevant markets are widely recognized as the origin-and-destination pair, thus concentration when appropriately calculated will be actually significantly higher than the values just presented. Because the service is operated under public delegation, through permits and authorizations, most routes are served by a single operator. Competition is allowed in a few routes, notably the most important ones, including the shuttle service between Rio de Janeiro-São Paulo, which carried 0.7 million passengers a year each way. The latter has competition in prices through discounts below the price cap and also in service attributes, as the route carries many coach travelers and may have an element of price differentiation; ii) there are major economic groups that own several operators in this industry so that actual concentration may be also higher for this reason; and iii) there is also a factor affecting concentration that is usually not accounted for, namely the existence of a growing informal market, so that concentration may be probably lower than suggested by calculations using official data, even when the analyst carefully considers the economic grouping by capital ownership and the appropriate relevant markets[6]. However, estimating the actual concentration in relevant markets is beyond the scope of this study.

The ownership structure is changing rapidly over recent years. Some big merger or acquisition deals are taking place with relevant impact on market structure. It is also noteworthy that the Gol group has its origins in the coach industry and keeps growing by acquisitions in both the coach and airline business. In the latter, Gol recently incorporated Varig, which was formerly the leading national airline.

The regulatory framework in the coach transportation industry has considerably changed since 2000. Regulation of interstate and international lines was transferred in 2002 to ANTT - Agência Nacional de Transportes Terrestres[7]. Scheduled routes are now granted by ANTT through competitive bidding, in a process that requires the signature of a contract of concession. The operation of charter coach services is subject to a more swift process of authorization. ANTT is taking further steps to enhance competition in markets with more density and is also trying to increase transparency by publishing relevant information such as tariffs in their website. So far the user can get to know the price cap for any route, its frequencies, special attributes offered and some additional information. However, she will not be able to find out whether an operator is offering discounts below the price cap. In 2007, a quasi-deregulation of prices was implemented by the Resolution number 1.928/07 of ANTT. Within this new regulatory framework, carriers can now have price discrimination within the same coach and therefore can put into practice the same yield management tactics extensively employed by the airlines. The price cap system is still a biding restriction, however.

## III. Modeling Competition Between Alternative Transportation Models

In this section we develop a theoretical framework of competition between transportation modes that can be empirically handled for a test of intermodal interaction. Consider the demand of transportation mode $i$ at time $t$. Assume intermodal competition between coaches ($i = 1$) and airlines ($i = 2$), and the following demand system:

$$
\begin{cases}
q_{1t} = \alpha_1 p_{1t}^{-\beta_{11}} p_{2t}^{\beta_{12}} Y_t^{\delta_1} \\
q_{2t} = \alpha_2 p_{1t}^{\beta_{21}} p_{2t}^{-\beta_{22}} Y_t^{\delta_2}
\end{cases}
\tag{1}
$$

where $p_{it}$ and $q_{it}$ are, respectively, the average fare and total quantities of mode $i$ at time $t$. $\alpha_i$ are parameters indicative of "absolute advantage in demand" (Dixit, 1979); and $\beta_{ii}$ and $\beta_{ij}$ are, respectively, mode-specific own and cross price elasticities of demand. $Y_t$ is a proxy for income and levels of economic activity and $\delta_i$ is the resulting income elasticity. $\alpha_1$, $\alpha_2$, $\beta_{ij}$, $\delta_{ij} > 0$, i, j = {1, 2}.

The costs side is modeled in the following way:

$$
TC_{it} = FC_{it} + \frac{\varphi_{it}}{2} q_{it}^2, \quad FC_{it} > 0
\tag{2}
$$

Where $TC_{it}$ is the total cost of mode $i$ at time $t$ and $FC_{it}$ and $\varphi_{it}$ are mode-specific, time-varying, costs parameters. $FC_{it}$ is the fixed cost term and $1/\varphi_{it}$ represents returns to density of mode $i$ at time $t$. Marginal cost and total profits of mode $i$ at time $t$ are, respectively, $c_{it} = \varphi_{it} q_{it}$ and $\pi_{it} = p_{it} q_{it} - TC_{it}$.

Assume price competition, which implies that modes are static Bertrand opponents. First-order condition for profit maximization (*FOC*) of mode $i$ at time $t$ is then the following:

$$
\frac{d\pi_{it}}{dp_{it}} = 0 \quad \longleftrightarrow \quad q_{it} + q'_{it} p_{it} - q'_{it} c_{it} = 0 \quad \longleftrightarrow \quad p_{it} = c_{it} - \frac{q_{it}}{q'_{it}}
\tag{3}
$$

where $q'_{it} = dq_{it}/dp_{it}$. With some additional manipulation in (3), one has the following *FOC* of mode $i$:

$$p_{it} = \varphi_{it} q_{it} - \frac{\alpha_i p_{it}^{-\beta_{ii}} p_{jt}^{\beta_{ij}} Y_t^{\delta_i}}{\left(-\beta_{ii} p_{it}^{-1}\right) \alpha_i p_{it}^{-\beta_{ii}} p_{jt}^{\beta_{ij}} Y_t^{\delta_i}} = \varphi_{it} q_{it} + \frac{p_{it}}{\beta_{ii}} \tag{4}$$

And finally the *FOC* becomes:

$$p_{it} = \eta_i \varphi_{it} q_{it} \tag{5}$$

where $\eta_i = \beta_{ii}/(\beta_{ii}-1)$. By plugging (1) into (5) and then manipulating terms one has:

$$p_{it}^{1+\beta_{ii}} = \alpha_i \eta_i \varphi_{it} Y_t^{\delta_i} p_{jt}^{\beta_{ij}} \tag{6}$$

We can therefore reach the best-curve response (reaction function) of mode $i$ with respect to rival mode $j$:

$$p_{it} = R_i(p_{jt}) = \left(\alpha_i \eta_i \varphi_{it} Y_t^{\delta_i}\right)^{\mu_i} p_{jt}^{\mu_i \beta_{ij}} \tag{7}$$

where $\mu_i = 1/(1+\beta_i)$ and $R_i(p_{jt})$ is the reaction function of mode $i$. By considering a system of equations $R = R_i(p_{jt})$, $(i, j = 1, 2)$ it is possible to inspect the existence and unicity of a Bertrand-Nash equilibrium, which would constitute the solution of (7). Under regularity conditions on the parameters in (1) and (2), a unique Nash equilibrium in prices will emerge.

Equation (7) also permits the researcher to infer the consequences from demand and supply shocks in this situation of market equilibrium under intermodal competition. Figure 2 indicates the effects of exogenous changes in the parameters and the resulting Bertrand-Nash equilibria. Suppose the market is initially in F. It is clear that a cost shock (an increase in road tolls or in the unit cost of diesel, for example) would provoke an upward shift in mode 1's reaction curve from $R_1^0(p_2)$ to $R_1^1(p_2)$, increasing prices of both modes to $p_1^1$ and $p_2^1$, and a new Bertrand-Nash equilibrium in $G$.

Now suppose the entry of a low cost airline. A change in market structure of transportation mode 2 (airlines) would cause both curves to become flatter, as a result of the increase in the perceived rivalry in the market –by either a decrease in $\beta_{ii}$ or an increase in $\beta_{ij}$, or both. There would be new reaction functions $R_1''(p_2)$ and $R_2''(p_1)$ now. The resulting intermodal competition effect would be a change from $F$ to a new Bertrand-Nash equilibrium in $H$, in which prices of both transportation modes are lower ($p_1''$ and $p_2''$).

In order to econometrically handle expression (7) we will first perform its linearization by logarithms:

$$\ln p_{it} = \mu_i \ln \alpha_i + \mu_i \ln \eta_i + \mu_i \ln \varphi_{it} + \mu_i \delta_i \ln Y_t + \mu_i \beta_{ij} \ln p_{jt} \tag{8}$$

Secondly, extract its first-difference:

## FIGURE 2

### THE EFFECTS OF SHOCKS IN COSTS AND MARKET STRUCTURE
ON EQUILIBRIUM FARES[8]

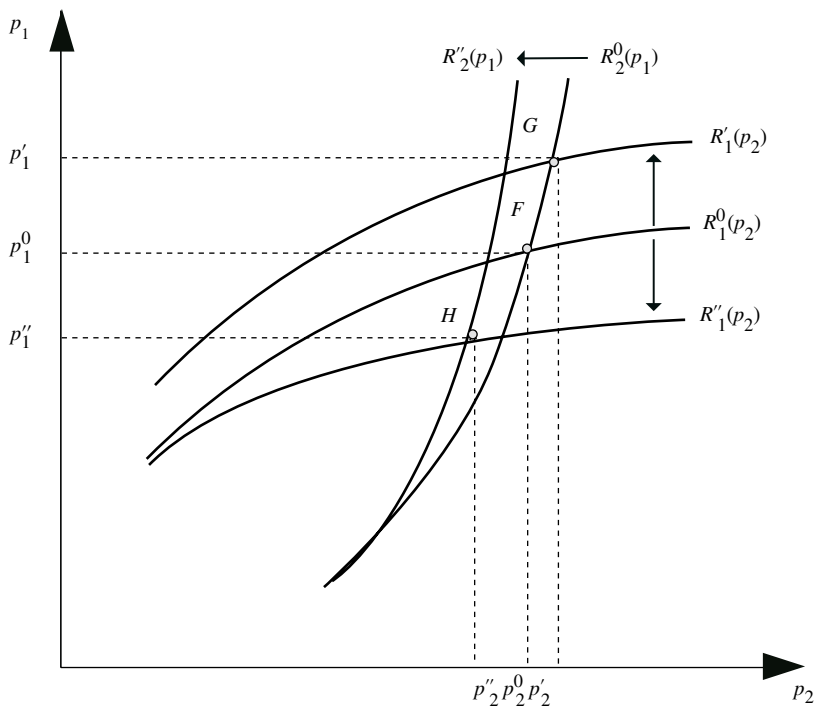

$$\Delta \ln p_{it} = \ln p_{it} - \ln p_{it-1} = \mu_i \Delta \ln \varphi_{it} + \mu_i \delta_i \Delta \ln Y_t + \mu_i \beta_{ij} \Delta \ln p_{jt} \qquad (9)$$

And finally, let us propose the following specification of cost parameter $\varphi_{it}$:

$$\Delta \ln \varphi_{it} = \varphi_0 + \sum_h \varphi_{hi} \Delta \ln W_{hit} \qquad (10)$$

where $W_{ht}$ is a vector of cost shifters of $\varphi_{it}$. By inserting (10) into (9):

$$\Delta \ln p_{it} = \mu_i \varphi_0 + \sum_h \mu_i \varphi_{hi} \Delta \ln W_{hit} + \mu_i \delta_i \Delta \ln Y_t + \mu_i \beta_{ij} \Delta \ln p_{jt} \qquad (11)$$

We then have the final empirical specification of the pricing equation of trans-
portation alternatives within a framework of intermodal competition:

$$\Delta \ln p_{it} = \kappa_{1i} + \sum_h \kappa_{2hi} \Delta \ln W_{ht} + \kappa_3 \Delta \ln Y_t + \kappa_4 \Delta \ln p_{jt} \qquad (12)$$

where $i = \{1,2\}$ and $k_{1i}$, $k_{2hi}$, $k_3$ and $k_4$ are parameters of the reaction function.

## IV.  The Empirical Approach

Our main objective here is to make inference about coach operators' pricing behavior and above all to infer the significance of intermodal competition in Brazil from available data. We perform this empirical test by estimating the sensitivity of changes in coach fares to changes in air travel fares once cost and demand observables and unobservables are controlled for.

The dataset used for the empirical model was collected from Instituto Basileiro de Geografia e Estatística (IBGE), which produces the National System of Consumer Price Indexes; more precisely, we used the same data on transportation prices change collected for the Extensive National Consumer Price Index - IPCA. It is basically a set of monthly series from August 1999 to November 2005, disaggregated for the following geographic areas in Brazil: metropolitan regions of Belém, Fortaleza, Recife, Salvador, Belo Horizonte, Rio de Janeiro, São Paulo, Curitiba and Porto Alegre, Brasília and municipal district of Goiânia. We had to drop Belém, Fortaleza and Recife, however, as we detected no series of interstate travel price change in the original data, as obtained from one of IBGE's websites[9]. Therefore, the final sample size had 525 observations ($j = 7$ regions and $t = 75$ months), in a balanced panel data.

One caveat about the data used here: as the most disaggregated unit of observation available from IBGE´s dataset was measured at the city-level, one has to be cautious when interpreting estimation results, as it is well recognized that the relevant market in transportation is actually the origin-and-destination pair. Therefore, our city fares analysis must be interpreted as the study of average behavior across origin-and-destination pairs. Besides that, we cannot observe average stage length for both coaches and airlines.

In other to develop the empirics, here we follow the airline literature such as Borenstein (1989) and Evans and Kessides (1993). We therefore have an empirical model of pricing decisions of firms in the market. The empirical framework, consistent with the reaction function of (12), is the following:

$$\Delta coach_{jt} = k_1 \Delta diesel_{jt} + k_2 \Delta tire_{jt} + k_3 \Delta toll_{jt}$$
$$+ k_4 \Delta airline_{jt} + \gamma_j + \gamma_t + \varepsilon_{jt} \qquad (13)$$

Where the mathematical operator $\Delta$ here means price change (*in percentage*) within one unit of time, that is, $\Delta X_t$ is approximately equal to ($\ln X_t - \ln X_{t-1}$).

$\Delta coach_{jt}$ is the average change in interstate coach fare in city $k$ and time $t$; $\Delta diesel_{jt}$, $\Delta tire_{jt}$ and $\Delta toll_{jt}$ are cost shifters, representing, respectively, the average change in the unit prices of diesel, tire, and tolls in city $j$ and time $t$; $\Delta airline_{jt}$ is the average change in the price of air travel in city $j$ and time $t$. All variables are expressed in real terms, that is, once discounted each period's inflation rate for the consumer (IPCA index).

$\gamma_j$ and $\gamma_t$ are, respectively, city-specific and time-specific effects, designed to control for potential unobservables across cities and time in a two-way panel. They also account for national time-varying income effects that are common across cities, and for income asymmetries between cities that do not change across time. Also, they control for time-invariant components of the choice of passengers between coaches and self transportation in the local markets. $\varepsilon_{jt}$ is the error term and $\kappa_1,..., \kappa_4$ are the unknown parameters. It is important to emphasize that, as seen in Section III, the parameter $\kappa_4 = \mu_i\beta_{ij} = \beta_{ij}/(1+\beta_i)$ has an important theoretical meaning: it relates the own and cross price elasticities of the demand for the modes of transportation. Also, as road transportation may be considered an essential good for Brazilian travelers, $\beta_i$ tends to be small; therefore, although being an unobservable, $\kappa_4$ tends to assume a value that is close to $\beta_{ij}$ (the cross-price elasticity of demand) and hence may represent a rough approximation of the actual substitution patterns by consumers in the market.

Below is Table 1, with some descriptive statistics for the variables in (13); note that, for simplicity, indexes k and t are omitted from now on:

TABLE 1

DESCRIPTIVE STATISTICS[10]

| Variables | Mean | Std. Dev. | Min. | Max. |
|---|---|---|---|---|
| $\Delta coach$ | 0.212 | 3.299 | −8.060 | 15.210 |
| $\Delta Diesel$ | 0.713 | 2.628 | −6.710 | 10.560 |
| $\Delta Tire$ | 0.256 | 1.990 | −6.690 | 7.000 |
| $\Delta Toll$ | 0.913 | 2.887 | −3.520 | 18.970 |
| $\Delta Airline$ | 0.769 | 4.193 | −19.810 | 24.440 |

We treat $\Delta airline_{kt}$ as jointly determined with $\Delta coach_{kt}$. In order to account for this endogeneity of regressor $\Delta airline_{kt}$ we employ simultaneous equations methods. The estimator used here is the two-step efficient generalized method of moments (GMM) estimator, and generates coefficient estimates that are efficient in the presence of both arbitrary heteroskedasticity and arbitrary autocorrelation of first-order. Also, as (13) permits observing, there are both times (months) and section (cities) fixed effects.

In order to properly identify the effects of $\Delta airline_{kt}$, in (13), we need instrumental variables that are valid and relevant. By validity we mean orthogonal to the error term and by relevance we mean correlated with $\Delta airline_{kt}$. Our methodology is similar to the one proposed by Hausman, Leonard and Zona (1994) to create instruments: we then use airline fare changes in the other remaining cities to instrument fare changes in a given city. Additionally, we make use of tests of validity (Hansen J test) and relevance (Anderson canonical correlations likelihood-ratio test) to check whether the instruments are adequate. In both tests, the adequacy of the proposed set
of instruments is confirmed.

Table 2 presents the estimation results.

TABLE 2

ESTIMATION RESULTS[11]

| Variables | $\Delta$ Coach |
|---|---|
| Constant | 3.186‡ |
| | (1.131) |
| $\Delta$ Diesel | 0.267‡ |
| | (0.086) |
| $\Delta$ Tire | 0.255‡ |
| | (0.052) |
| $\Delta$ Toll | 0.516‡ |
| | (0.136) |
| $\Delta$ Airline | 0.311‡ |
| | (0.105) |
| city - Brasília | −0.201 |
| | (0.193) |
| city - Curitiba | 0.056 |
| | (0.165) |
| city - Goiânia | 0.044 |
| | (0.216) |
| city - Rio de Janeiro | 0.234 |
| | (0.208) |
| city - Salvador | 0.156 |
| | (0.235) |
| city - São Paulo | −0.118 |
| | (0.170) |
| Adjusted $R^2$ | 0.833 |
| MSE | 1.345 |
| F Statistic | 30.310‡ |
| Anderson Statistic | 50.360‡ |
| Hansen Statistic | 1.390 |
| Number of Observations | 525 |

‡ Significant at 1% level.

With Table 2, it is possible to observe that the coefficients of the cost shifters
$\Delta diesel_{kt}$, $\Delta tire_{kt}$ and $\Delta toll_{kt}$ are statistically significant and have the signs in accor-
dance with theory. $\Delta toll_{kt}$ has the biggest effect on prices, which is possibly due to
the strong correlation between the higher tolls and the higher quality of service of
the road infrastructure in São Paulo. In fact, this result of $\Delta toll_{kt}$ may be indicative
that coaches make use of the better highways and facilities of the state to charge
travelers more than proportionally and may be suggestive that the regulator should be concerned with the entry barriers at the highly congested Tietê Terminal in São Paulo – the second largest coach terminal in the world.

The pricing model estimated and presented in Table 2 clearly provides some evidence on the degree of intermodal competition that characterizes the airline and coach industries in Brazil. We tested for the presence of intermodal interaction (probably due to a substitution effect) among air and road transportation for the interstate market, and found a highly significant effect. In fact, one can reject, at one percent of significance, the null hypothesis that airline fare variations have no effect on coach fare variations. This result is achieved once relevant cost components are accounted for – that is, price variation of diesel, tire and toll. This is quite a relevant outcome because the coach travel industry is usually regarded as a major cartel in Brazil, that is, not subject to internal competitive pressure; thus, the results here indicate that there is significant competition stemming from intermodal competition with airlines[12].

It is important to emphasize that the interaction between the air and road markets for passenger transportation may take place in their borders. *Premium coach passengers*, that is, those who are able to pay for the highest fares as they have higher sensitivity to coach quality, are the effective *marginal consumers* in this market; therefore they constitute the small but increasing set of passengers who consider changing means of transportation when price differences narrow down. So the airline industry has been from time to time appealing to the upper-class coach passengers, who go for flying during air fare wars or deep discount periods, and eventually go back to the coach service in ordinary periods. The estimated coefficient of $\Delta airline_{kt}$, presented in Table 2, suggests that cross-elasticities between air and ground transportation of passengers are no longer negligible in Brazil but, on the contrary, it is indicative of a situation of strong and relevant intermodal interaction. This result has interesting implications for competition analysis since it raises the issue on whether the relevant market continue to be defined as the individual mode, both for airline and coach cases.

## V. Conclusions

The objective of this paper was to identify and measure potential intermodal competition in the interstate travel market in Brazil. By estimating a reduced-form model of pricing decisions of coach operators, it was possible to test whether the intermodal interaction between air and road transportation was significant.

Results indicated that coach fares are sensitive to air fares, indicating a very significant interaction and potential substitution effect. This is quite a relevant outcome because the coach travel industry is usually regarded as a major cartel in Brazil and therefore not usually subject competitive pressure. The most probable explanation for that is related to the presence of premium coach passengers, with higher willingness-to-pay, who go for flying during air fare wars or deep discount periods, and eventually go back to the coach service in ordinary periods. It is noteworthy that during price war periods long distance air tickets may be priced below coach prices. Besides, some services are being upgraded by some coaches and some are being reduced by airlines. In some sense, this may reflect a sort of "convergence process" between means of passenger transportation in some relevant markets for premium passengers, analogously to what happens in telecommunications although in a much less important degree.

In addition, the result challenges the traditional way of defining relevant markets for passenger transportation competition in Brazil. The positive cross-elasticities may suggest that the product scope of relevant markets in some origin-and-destination pairs may be wider than the single mode that is often considered, or at least that the analyst should pay closer attention to the intermodal interactions. This may be true not only for the interaction between air transportation and coaches, but also for the possibilities involving railroads and waterways in some very specific markets inside Brazil.

The empirical analysis performed here also served as a test of the efficacy of the recent liberalization measures in both airline and coach industries in inducing more efficient and competitive transportation and logistics in Brazil.

## Notes

[1]  Source: 2001 Statistical Yearbook of Transportation of Brazil, available at the website www.geipot. gov.br/anuario2001/rodoviario/rodo.htm.
[2]  Source: IBGE.
[3]  The United States possesses the most extensive domestic highway system in the world (source: website www.estradas.com.br)
[4]  Urban rail transportation is by far more significant, with 726 million carried in 2006 (ANTT, 2007)
[5]  Concentration ratio for the top four firms in terms of passengers times kilometers.
[6]  Some analysts estimate that nowadays the informal market may account for almost fifteen percent of the total market in Brazil.
[7]  States regulate their intra-state routes while munnicipalities regulate their local services.
[8]  Note that index $t$ is omitted.
[9]  The IBGE aggregated data bank available at http://www.sidra.ibge.gov.br/. The IPCA series numbers were: 5101007 (coach), 5104003 (diesel), 5102010 (tire), 5102015 (toll) and 5101010 (airline).
[10] Variables expressed in nominal percentage.
[11] Estimated time-specific effects ($\gamma_t$) omitted. Belo Horizonte city is the base-case of the set of city dummies.
[12] Note that the city effects were not statistically significant, meaning that there is no local-specific components of variations in coach fares.